# NURE: an ERC project to "measure" the nuclear matrix elements

**Manuela Cavallaro**[1]
*INFN – Laboratori Nazionali del Sud*
*Via S. Sofia 62, 95125, Catania, Italy*
*E-mail:* manuela.cavallaro@lns.infn.it

The NURE (NUclear REactions for neutrinoless double beta decay) project has been selected for receiving funding in the call Starting Grant 2016 of European Research Council (ERC). The project, which takes advantage of nuclear physics methods to give a contribution to the neutrino physics, has a duration of 5 years. A key aspect is the use of the K800 Superconducting Cyclotron, for the acceleration of the required high resolution and low emittance heavy-ion beams, and of the MAGNEX large acceptance magnetic spectrometer, for the measurement of the reaction cross sections. These facilities are in operation at INFN Laboratory Nazionali del Sud in Catania (Italy).

---

[1]Speaker

Introduction

The research on neutrinoless double beta decay, 0νββ, is at the present time strongly pursued both experimentally and theoretically. The observation of this phenomenon will determine whether neutrino is a Dirac or Majorana particle and will provide a measurement of the absolute neutrino mass scale, which is nowadays one of the fundamental problems in physics.

Since the 0νββ decay is a nuclear transition, its analysis necessarily implies nuclear structure issues and nuclear physics plays a key role in this field. The nuclear matrix elements (NME) connecting the initial and final wave functions of the nuclear states involved in the decay enter in the expression relating the 0νββ decay rate to the average neutrino mass. Thus the knowledge of NME gives access to the neutrino mass, if the 0νββ decay rate is measured.

The ERC project NURE plans to use double charge-exchange (DCE) nuclear reactions to extract information on 0νββ NME. The basic point is that the initial and final state wave-functions are the same in the two classes of processes (0νββ and DCE) and the transition operators are similar, i.e. in both cases present a superposition of Fermi, Gamow-Teller and rank-two tensor components with a relevant momentum available. Thus, even if the two processes are mediated by different forces, the NME are connected and the measurement of DCE cross-sections can give crucial information on ββ NME.

NURE plans to carry out a campaign of experiments at INFN-LNS using accelerated ion beams on two targets candidates for 0νββ decay. Due to the very low cross sections of DCE reactions and to the limitation of the total beam time available, it is prohibitive to extend this research to all the systems candidate for 0νββ decay (about a dozen), with the present set-up. In this respect NURE has been conceived and will develop in a wider context, within an international collaboration, the NUMEN project [1], [2], [3]. An intense activity of R&D to increase the high-rate tolerance of detectors and targets will be conducted in the next years by the NUMEN collaboration. Also the theoretical aspects, crucial to find the connections between double charge exchange reaction cross sections and 0νββ nuclear matrix elements, will be developed.

**1. Double charge exchange reactions**

Double charge-exchange reactions (DCE) are processes characterized by the transfer of two units of the isospin component (two protons transformed into two neutrons or vice versa), leaving the mass number unchanged. The initial and final nuclear states involved in DCE reaction and *ββ* decay are the same and the transfer operators have similar spin-isospin mathematical structure. Namely they both contain a Fermi, a Gamow-Teller and a rank-two tensor term. A relevant amount of linear momentum (of the order of 100 MeV/c) is available in the virtual intermediate channel in both processes. This is an important similarity since the nuclear matrix elements strongly depend on the momentum transfer and other processes (single charge-exchange reactions, 2νββ decay etc.) cannot probe this feature. Thus, the connection between the involved nuclear matrix elements deserves a deep study and comprehension.

One should remind that a similar link is well established at a level of few percent between single *β* decay strengths and single charge-exchange reaction cross-sections, under specific dynamical conditions. Indeed, single charge-exchange reactions are routinely used as a tool to determine Fermi and Gamow-Teller transition strengths for single *β* decay, as demonstrated by several papers and reports [4]. However, no one has ever tested this proportionality between *ββ*- decay strength and DCE cross-section.

Experimental attempts were done in the past to perform DCE reactions [5], [6]. However, most of them were not conclusive because of the very poor yields in the measured energy spectra and the lack of angular distributions, due to the very low cross-sections involved. High resolution spectra and angular distributions are in fact crucial to identify the transitions of interest and separate the direct reaction mechanism among other possible multi-step processes that can hide the desired information [7], [8].

The advent of new facility based by large acceptance and high resolution spectrometers have allowed to obtain very promising results in this field. One should mention for example the recent measurements on $^{48}$Ca($^{12}$C,$^{12}$Be)$^{48}$Ti and $^{12}$C($^{18}$O,$^{18}$Ne)$^{12}$Be [9] and the results of the experiment at INFN-LNS on $^{40}$Ca($^{18}$O,$^{18}$Ne)$^{40}$Ar [10]. This latter has shown that high resolution energy spectra and angular distribution can be accessed and matrix elements extracted within a promising level of accuracy at least for the $^{40}$Ca case.

## 2. The NURE project

Nowadays DCE reactions induced by heavy-ion beams can be studied with high accuracy and resolution in very few specialized laboratories worldwide, such as the Istituto Nazionale di Fisica Nucleare - Laboratori Nazionali del Sud (INFN - LNS). There, the Superconducting Cyclotron (SC) accelerator is available to delivery several beams in a broad energy range, and it is possible to detect the reaction ejectiles using the MAGNEX large acceptance magnetic spectrometer [11], [12], [13].

NURE plans to carry out a campaign of experiments by using heavy-ion accelerated beams impinging on two targets of interest as candidate nuclei for the $\beta\beta$ decay. The DCE channel will be explored by the ($^{18}$O,$^{18}$Ne) reaction for the $\beta^+\beta^+$ direction in the target and the ($^{20}$Ne,$^{20}$O) for the $\beta^-\beta^-$. Moreover, the complete net involving the multi-step transfer processes characterized by the same initial and final nuclei will be also studied in the same experimental setup. In particular the ($^{18}$O,$^{16}$O) and ($^{20}$Ne,$^{22}$Ne) two-neutron transfer, ($^{18}$O,$^{20}$Ne) and ($^{20}$Ne,$^{18}$O) two-proton transfer, ($^{18}$O,$^{18}$F) and ($^{20}$Ne,$^{20}$F) single charge-exchange will be measured. An example of the net that will be explored for one of the targets of interest ($^{76}$Ge) is sketched in figure 1.

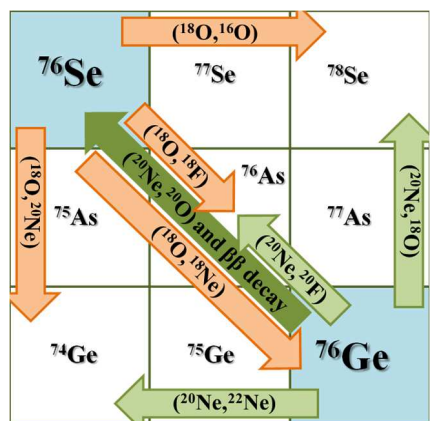

Figure 1. Scheme of the complete net of processes that will be studied in the case of the $^{76}$Ge–$^{76}$Se pair of nuclei of interest for the $\beta\beta$ decay. Inside the arrows the reaction which populates the final nuclei is shown.

The SC accelerator provides all the required beams for the project proposed here, namely $^{18}$O and $^{20}$Ne at energies ranging from 10 MeV/u to 60 MeV/u with an excellent energy resolution and emittance [14].

MAGNEX was designed to investigate processes characterized by very low yields. It allows the identification of heavy ions with high mass (~ 1/200), angle (~ 0.2°) and energy resolution (~ 1/1000), within a large solid angle (~ 50 msr) and momentum range (~ 25%) [15-20]. It also allows to measure at zero degrees, which is the most important region for this research, thus creating ideal laboratory conditions for the present proposal. High resolution measurements for other reactions characterized by cross-sections falling down to tens of nb/sr were already performed by this setup [21]. The crucial issue is the implementation of the powerful technique of trajectory reconstruction, which allows to solve the equation of motion of each detected particle to very high order (10$^{th}$ order). In this way an effective compensation of the high order aberrations induced by the large aperture of the magnetic elements is achieved [22-25]. The use of the sophisticated data reduction approaches based on the differential algebra is a unique feature of MAGNEX, developed in the years. This guarantees the above mentioned performances and its relevance in the worldwide research of heavy-ion physics.

Two pairs of target nuclei of interest as candidate for *0νββ* will be studied in both *β⁻β⁻* and *β⁺β⁺* direction: the $^{76}$Ge-$^{76}$Se and the $^{116}$Cd-$^{116}$Sn pair. These targets represent candidates for *0νββ* decay and belong to two different classes of nuclei: those in which protons and neutrons occupy the same major shells ($^{76}$Ge-$^{76}$Se) and those in which they occupy different major shells ($^{116}$Cd-$^{116}$Sn). Moreover, the magnitude of the Fermi matrix element, which is related to the overlap of the proton and neutron wave functions, is different in these two classes of nuclei, being large in the former and small in the latter case. The study of the different behaviour of the nuclear matrix elements in these systems is interesting. These nuclei have been chosen because the transition to the first excited states can be well separated from the ground states by the MAGNEX resolution (being the first excited states at 562 keV for $^{76}$Ge, 559 keV for $^{76}$Se, 1.29 MeV for $^{116}$Sn and 513 keV for $^{116}$Cd) and also because the production technologies of these targets are already available at LNS.

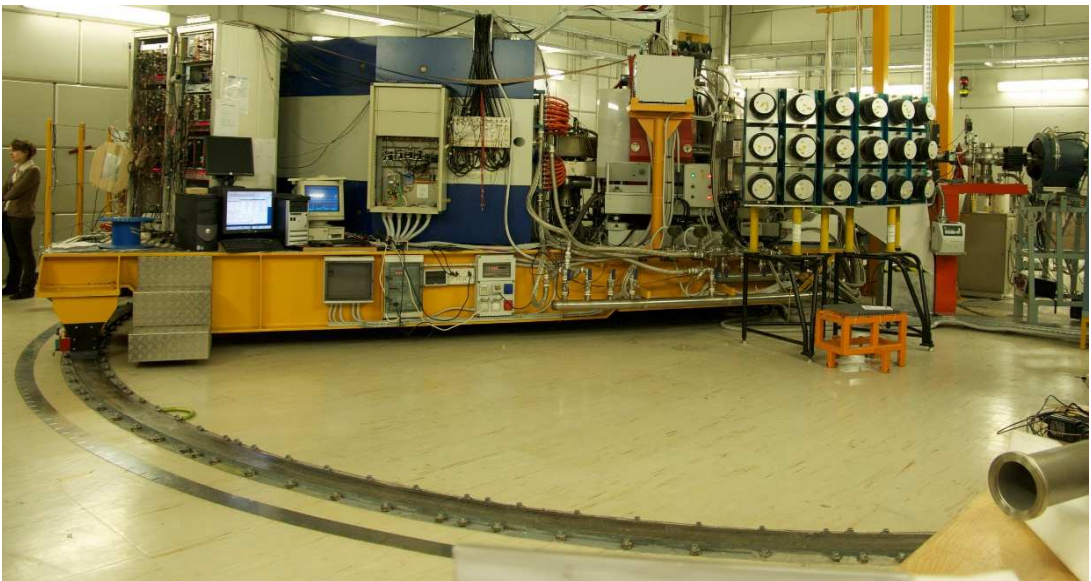

Figure 2. The MAGNEX spectrometer at INFN-Laboratori Nazionali del Sud

**Acknowledges**

This project has received funding from the European Research Council (ERC) under the European Union's Horizon 2020 research and innovation programme (grant agreement No 714625)